\documentclass{article}

\usepackage[preprint]{neurips_2022}

\usepackage[utf8]{inputenc} 
\usepackage[T1]{fontenc}    
\usepackage{hyperref}       
\usepackage{url}            
\usepackage{booktabs}       
\usepackage{amsfonts}       
\usepackage{nicefrac}       
\usepackage{microtype}      
\usepackage{xcolor}         
\usepackage[]{algorithmic} 
\usepackage{algorithm}
\usepackage{multirow}
\usepackage{amsmath} 
\usepackage{graphicx}

\newtheorem{problem}{Problem}
\newtheorem{remark}{Remark}
\numberwithin{remark}{section}

\newcommand{\Var}{\mathrm{Var}}

\newcommand{\EE}{\mathbb{E}}

\newcommand{\PP}{\mathbb{P}}

\newcommand{\RR}{\mathbb{R}}

\title{Deep Learning for Systemic Risk Measures}

\author{%
 {Yichen Feng \thanks{Equal contribution.}\qquad Ming Min \footnotemark[1] \qquad Jean Pierre Fouque \thanks{Work supported by NSF grants DMS-1814091 and DMS-1953035.}}  \\
  Department of Statistics and Applied Probability\\
  University of California, Santa Barbara\\
  Santa Barbara, CA 93106 \\
  \texttt{feng@pstat.ucsb.edu, m\_min@pstat.ucsb.edu} \\
  \texttt{fouque@pstat.ucsb.edu} \\
}

\begin{document}

\maketitle

\begin{abstract}
  The aim of this paper is to study a new methodological framework for systemic risk measures by applying deep learning method as a tool to compute the optimal strategy of capital allocations. 
Under this new framework, systemic risk measures can be interpreted as the minimal amount of cash that secures the aggregated system by allocating capital to the single institutions \textit{before aggregating} the individual risks. This problem has no explicit solution except in very limited situations. Deep learning is increasingly receiving attention in financial modelings and risk management and we propose our deep learning based algorithms to solve both the primal and dual problems of the risk measures, and thus to learn the \textit{fair} risk allocations. In particular, our method for the dual problem involves the training philosophy inspired by the well-known Generative Adversarial Networks (GAN) approach and a newly
designed direct estimation of Radon-Nikodym derivative. We close the paper with substantial numerical studies of the subject and provide interpretations of the risk allocations associated to the systemic risk measures.
In the particular case of exponential preferences, numerical experiments demonstrate excellent performance of the proposed algorithm, when compared with the optimal explicit solution as a benchmark.
\end{abstract}

\section{Introduction}
\paragraph{Systemic Risk Measures.} Traditional risk management strategies of financial systems, which mainly treated individual institutions as isolated, insufficiently capture the systemic risk that is generated by connections among system components and their contagion. The outburst of the global financial crisis has dramatically brought awareness of the urgent need for novel approaches that capture systemic riskiness with interconnected financial components, such as institutions and members of clearing houses. 

The classical conceptual framework of univariate monetary risk measures was first introduced by the seminal paper of Artzner et al. \cite{artzner1999coherent} and over the last few years, the literature on systemic risk has been growing fast taking in account different points of view on the subject.
Properties of measures and duality were emphasized in the studies of F{\"o}llmer and Schied \cite{follmer2002convex} and Frittelli and Rosazza Gianin \cite{frittelli2002putting}. Systemic risks characterized by interbank lending  has been studied via mean field approach in Fouque and Sun \cite{sun2014systemic}, Fouque and Ichiba \cite{fouque2013stability}, Carmona et al. \cite{carmona2013mean}. We refer the reader to Armenti and Cr{\'e}pey \cite{armenti2017central} for deterministic systemic risk allocations and the solutions to the classical optimization problem. For empirical studies on the default fund of Central Counterparty Clearing houses, we refer for example to Feng et al. \cite{feng2022systemic}, Armenti et al. \cite{armenti2018multivariate}, and Ben-Tal and Teboulle \cite{ben2007old}. See also the volume of Fouque and Langsam \cite{fouque2013handbook} for an exhaustive overview on the literature on systemic risk.

Methodological frameworks based on acceptable allocations have been proposed by several studies. Biagini and Fouque et al. \cite{biagini2019unified} generalize univariate framework to a multivariate setting where systemic risk measures are designed via multivariate acceptance sets which is very flexible and comprises many of the existing systemic risk measures proposed in the literature. While they mainly conduct systematic studies of the properties of the proposed measures in terms of set valued functions, diversification and monotonicity in \cite{biagini2019unified,BFFM-fairness}, the explicit solution can be found only in very limited cases.

\paragraph{Deep Learning.} 
Deep learning has been a hot topic in many fields that aims to establish an automatic algorithm to improve performance of tasks. The architectures of deep learning are conceptually inspired by the structure of the brain and have been applied to fields including computer vision, natural language processing and scientific computing. The financial sector is adopting deep learning techniques and thus entering a new era of rapidly developing. Deep learning shows great power and enables significant opportunities in financial modeling and risk management; see for instance Heaton et al. \cite{heaton2016deep} for financial prediction, Min and Hu \cite{min2021signatured}, Hu \cite{hu2019deep} and Cao et al. \cite{cao2020connecting} for solving stochastic differential games, Wise et al. \cite{wiese2019deep} for deep hedging, to list a few. 
A very popular subclass of deep neural networks is generative adversarial network (GAN) \cite{goodfellow2014generative} that contains two neural networks contest with each other in a game.
We refer the reader to Ni et al. \cite{ni2021siggan} for an alternative GAN method with application of sequential data generation. The optimization of GAN is to solve a min-max problem, which shed some lights on our algorithm design in this paper.

\paragraph{Contribution.} We design deep learning algorithms to study the systemic risk measures proposed in Biagini et al. \cite{BFFM-fairness} and improve it to a more realistic situation of scenario-dependent cash distribution without cross-subsidization. In our algorithms, we first learn the overall risk allocation of the system and the allocations to the individual financial components based on the primal problem. Then using a structure inspired by GAN, we solve an optimization problem over risk measures with a min-max objective function based on the dual representation of such systemic risk measures. One novelty of our paper is that, given input samples, we output Radon-Nikodym densities of new measure with respect to the original measure as random variables. 
It serves as a direct method to optimize over measure space and can be of independent interest as a pure data-driven approach for the financial math community. The optimizer in the dual formulation provides a risk allocation which is \textit{fair} from the point of view of the individual financial institutions. In Biagini et al. \cite{BFFM-fairness}, it turns out only in the case with exponential utilities, explicit solutions of overall risk allocations and \textit{fair} individual allocations of the proposed measures can be provided. However, to this end, our proposed algorithm can handle any utilities and is efficient for more realistic systemic risk measures.

\paragraph{Outline.}The structure of the paper is the following. In Section \ref{section2:setup}, starting from the multivariate risk measures, developed via multivariate acceptance sets in Biagini et al. \cite{biagini2019unified,BFFM-fairness}, we gradually introduce the dual representation of the measures and generalize the measures to be more realistic by adding nonnegativity to risk allocations. In Section \ref{section3:alg}, we focus on explaining the deep learning algorithms designed for the dual and primal problems. Section \ref{section4:exp} contains numerical examples and interpretations.

\section{Problem Setup}
\label{section2:setup}
To be self-contained, we provide a brief introduction of the systemic risk measures in this section. See Biagini et al. \cite{biagini2019unified,BFFM-fairness} for detailed definitions and more theoretical foundations.

Let $\mathcal{L}^0(\mathbb{R}^N)\, := \,\mathcal{L}^0(\Omega, \mathcal F ;\, \mathbb{R}^N)$ denote the space of $\mathbb{R}^N$-valued random variables on the probability space $\, (\Omega, \mathcal F , \mathbb P) \,$, and let $\mathbb{E}$ be the expectation under $\mathbb{P}$. In what follows, all needed integrability and regularity assumptions are taken for granted.
In the case of a system with $\,N\,$ interacting financial institutions, we assume a configuration of risky factors at a future time $T$ for this system is denoted by $\, {\mathbf X} \, :=\, (X^{1}, \ldots , X^{N})\in \mathcal{L}^0(\mathbb{R}^N)$.

Some systemic risk measures proposed in the literature (for example, \cite{artzner1999coherent, chen2013axiomatic}) are of the form
\begin{equation}\label{eqn:MeasureAfterAggragation}
    \rho(\mathbf{X})\,:=\,\inf\{m\in \mathbb{R}\,|\,\Lambda(\mathbf{X})+m\in \mathbb{A}\},
\end{equation}
where $\Lambda:\mathbb{R}^N\to\mathbb{R}$ is an aggregation function applied to the risk factor $\, {\mathbf X}\,$ in order to obtain a univariate random variable $\Lambda(\mathbf{X})$, and $\mathbb{A}\in\mathcal{L}^0(\mathbb{R}^N)$ is an acceptance set. 
When the univariate random variable $\Lambda(\mathbf{X})$ is interpreted as  some monetary loss, the classical scalar risk measure $\rho$ quantifies the minimal cash amount, that are needed today, to secure the system \textit{after aggregating individual risks} of the system, i.e. $\Lambda(\mathbf{X})$, possibly after a crisis has occurred. 

In contrast, to preventing crisis, another systemic risk measure can be defined of the form 
\begin{equation*}
    \rho(\mathbf{X}):=\inf\left\{\sum_{n=1}^N m_n\,|\,\mathbf{m}=\left(m_1,\ldots,m_N\right)\in \mathbb{R}^N\,,\, \Lambda(\mathbf{X}+\mathbf{m})\in \mathbb{A}\right\},
\end{equation*}
where allocations of additional capital $\mathbf{m}$ are added to $\mathbf{X}$ \textit{before aggregating risk components}. Furthermore, replacing the deterministic cash $\mathbf{m}$ with a random vector $\mathbf{Y}\in \mathcal{C}$ for some given class $\mathcal{C}$, choosing the aggregation function $\Lambda(\mathbf{x}) = \sum_{n=1}^N u_n(x_n)$
based on utility functions $u_n$, $n=1,\ldots,N$ of all financial institutions and taking the acceptance set 
$\mathbb{A} = \left\{Z\in L^1(\Omega,\mathcal{F};\mathbb{R}^N),\mathbb{E}[Z]\geq B\right\}$ for a given constant $B<0$
, we define our risk measure as follows.
\begin{problem}[Primal] The primal formulation of the systemic risk measure is given by
\begin{equation} 
\rho(\mathbf{X})\,:=\,\inf_{\mathbf{Y}\in \mathcal{C}}\left\{ \sum\limits_{n=1}^N Y^n \,|\, \mathbb{E}\left[\sum\limits_{n=1}^N u_n(X^n+Y^n) \right]\,\geq\, B \right\}.
\end{equation}
Here, $\mathcal{C}\subset \mathcal{C}_\mathbb{R}\cap M^{\Phi}$, where
\begin{align*}
&\mathcal{C}_\mathbb{R} := \left\{\mathbf{Y}\in L^0(\mathbb{R}^N)\,|\,\sum\limits_{n=1}^N Y^n\in \mathbb{R}\right\}, \quad M^{\Phi} : = M^{\phi_1}\times \cdots\times M^{\phi_N},
\end{align*}
and $M^{\phi_i}$ is the Orlicz heart, see Cheridito and Li \cite{cheridito2009risk}, corresponding to $\phi_n(x):=-u_n(-|x|)+u_n(0)$ for all $n$.
\label{problem:primal}
\end{problem}
Note that each $Y^n$, $n=1,\ldots,N$ is random and depends on the scenario $\omega$ realized at terminal time $T$, but the sum of random allocations is deterministic and known at the beginning, i.e. $\sum_{n=1}^N Y^n\in \mathbb{R}$. Thus the \textit{overall systemic risk} $\rho(\mathbf{X})$ can be interpreted as
the minimal total cash amount needed today to secure the system by distributing the cash at the future time $T$ among the components of the risk vector $\mathbf{X}$. 

However, it is also important to know how much each financial institution of the system contributes to the overall systemic risk. We denote the \textit{fair} risk allocation of each financial institution by $\rho^n(\mathbf{X})\in \mathbb{R}$ and it should satisfy the "Full Allocation" property, see Brunnermeier and Cheridito \cite{brunnermeier2019measuring} for example, which is $\sum_{n=1}^N\rho^n(\mathbf{X}) = \rho(\mathbf{X})$. According to Biagini et al. \cite{BFFM-fairness}, the solution is given by 
\begin{equation}\label{eqn:IndividualRiskAllocation}
    \rho^n(\mathbf{X})\, :=\, \mathbb{E} _{\mathbf{Q}^{n}_{\mathbf X}} [ Y^{n}_{\mathbf X} ]\,\quad \text{for}\quad n =1,\ldots,N,
\end{equation}
where $\mathbf{Q}_{\mathbf X}=(Q_{\mathbf X}^1,\ldots,Q_{\mathbf X}^N)$ is the optimizer of the dual problem stated below and the subscript ${\mathbf X}$ in \eqref{eqn:IndividualRiskAllocation} shows the dependence of $\mathbf{Q},\mathbf{Y}$ on the risk vector.

\begin{problem}[Dual] The dual representation of the systemic risk measure in Problem \ref{problem:primal} is given by
\begin{equation}
\rho(\mathbf{X}) = \max_{\mathbf{Q}\in \mathcal{D}}\left\{\sum_{n=1}^N\mathbb{E}_{Q^n}[-X^n]-\alpha_B(\mathbf{Q})\right\},\label{rho}
\end{equation}
where
\[ \alpha_B(\mathbf{Q}) = \sup_{\mathbf{Z}\in \mathcal{A}}\left\{ \sum_{n=1}^N\mathbb{E}_{Q^n}[-Z^n]\right\},
\]
with $
\mathcal{A} = \left\{\mathbf{Z}\in M^\Phi|\,\sum_{n=1}^N\mathbb{E}[u_n(Z^n)]\geq B\right\}\label{alpha}.
$
$\mathcal{D}$ is the domain where the probability measures $Q^n <<\mathbb{P}$ for all $n$ and the densities satisfy:

\begin{align}
\label{eq:dual_restriction_on_Q}
\bigg\{  \frac{d\mathbf{Q}}{d\mathbb{P}}\,\,\bigg|\,\, Q^n(\Omega)=1, \forall n \,\,\text{and}\,\, &\sum_{n=1}^N(\mathbb{E}_{Q^n}[Y^n]-Y^n)\leq 0 \nonumber \\
\text{for all }&\, \mathbf{Y}\in \mathcal{C}_\mathbb{R}\cap M^\Phi\bigg\}.
\end{align}

\label{problem:dual}
\end{problem}

The uniqueness and existence of both the individual optimal risk allocation vector $\mathbf{Y}$ in the primal Problem \ref{problem:primal} and the measure optimizer $\mathbf{Q}$ in the dual Problem \ref{problem:dual} are established in Biagini et al. \cite{BFFM-fairness}. Thus the optimal \textit{fair} risk allocation, defined as $\rho^n(\mathbf{X})\, =\, \mathbb{E} _{\mathbf{Q}^{n}_{\mathbf X}} [ Y^{n}_{\mathbf X} ]\in \mathbb{R}$ for all $n$ in \eqref{eqn:IndividualRiskAllocation}, also exists and it can be used as reference to collect cash at the beginning from each financial institution. Only in exponential utility case, there are explicit formulas for $\mathbf{Y},\mathbf{Q}$.

In the primal Problem \ref{problem:primal}, the choice of set $\mathcal{C}$ can affect the distributions of cash $\mathbf{Y}$. For example, choosing $\mathcal{C} = \mathbb{R}^N$ leads the result that $Y^n\in \mathbb{R}$ for all $n$, i.e. the allocation to each institution is determined today; for $\mathcal{C} = \mathcal{C}_\mathbb{R}$, the distribution can be chosen freely depending on the scenario $\omega$. But the latter case may result in negative cash allocations in some situations, which mean withdrawals of cash from certain components, and we call this phenomenon as \textit{cross-subsidization} between financial institutions. We then propose a more realistic systemic risk measure by requiring the individual risk allocation being nonnegative, based on the primal Problem \ref{problem:primal} without cross-subsidization.
\begin{problem}[No Cross-subsidization Systemic Risk Measure]
We add nonnegativity on the risk allocation $\mathbf{Y}$, and the new measure becomes
\begin{align}
\rho_+(\mathbf{X})\,:=\,\inf_{\mathbf{Y}\in \mathcal{C}_+}\left\{ \sum\limits_{n=1}^N Y^n \,|\, \mathbb{E}\left[\sum\limits_{n=1}^N u_n(X^n+Y^n) \right]\,\geq\, B \right\},
\end{align} 
where $\mathcal{C}_+\subset \mathcal{C}_\mathbb{R}^+\cap M^{\Phi}$ and
\[\mathcal{C}_\mathbb{R}^+:= \left\{\mathbf{Y}\in L^0(\mathbb{R}^N)\,|\,\sum\limits_{n=1}^N Y^n\in \mathbb{R},\, Y^n\geq 0 \,\forall \,n\right\}.\]
The definition of $M^{\Phi}$ is the same as the primal Problem \ref{problem:primal} and the dual formulation of the new measure is the same as defined in Problem \ref{problem:dual}, except that we need to add the condition $\mathbf{Z}\geq \mathbf{X}$ in the admissible set $\mathcal{A}$ and restrict the domain of $\mathbf{Y}$ in \eqref{eq:dual_restriction_on_Q} to be nonnegative, i.e. $\mathcal{C}_\mathbb{R}^+\cap M^{\Phi}$.
\label{problem:MoreRealisticSetup}
\end{problem}
The solutions of the no cross-subsidization setup, denoted by $\mathbf{Y}_+$ and $\mathbf{Q}_+$, doesn't have explicit solutions in any cases yet. We will present numerical solutions with interpretations in section \ref{section4:exp} to show the feasibility and superiority of our deep learning algorithm.

\section{Algorithms}
\label{section3:alg}
We state our deep learning algorithms for both primal problem and dual problem in this section. In particular, the dual problem involves optimization over the space of probability measures, we solve this problem by using a neural network to represent the Radon-Nikodym derivative. This idea may be of independent interest to mathematical finance community since it can be applied to achieve the change of measure from physical measure to risk neutral measure. 

\subsection{Primal Problem}
Denote by $\varphi: \mathbb{R}^N \to \mathbb{R}^N$ the fully connected neural networks parametrized by weights and biases $(w, b)$, that take the risk factor $\mathbf{X}(\omega)\in \mathbb{R}^N$ as input and generate cash allocation $\mathbf{Y}(\omega)\in \mathbb{R}^N$ as output for any scenario $\omega\in\Omega$. To ease the notation, we will omit $\omega$ in the rest of the paper when the context is clear. More precisely, 
$$\mathbf{Y} := (Y^1, \dots, Y^N) = (\varphi_{1}(\mathbf{X}), \dots, \varphi_{N}(\mathbf{X}) )=:\varphi(\mathbf{X}).$$

The primal Problem \ref{problem:primal} requires the total cash added to the system to be deterministic, we thus add a penalty for the variance of total cash allocation to the loss function. To deal with the acceptance set restriction in the objective, we add a second penalty for the failure cases.

For the primal problem, the objective function for our deep learning task becomes
\begin{align}
\label{eq:rho_tilde}
    J_{\text{primal}}(\varphi) := \sum_{i}& \varphi_i(\mathbf{X})  + \mu \cdot \Var\big(\sum_i \varphi_i(\mathbf{X})\big) \nonumber \\
    &+ \lambda \cdot \big(B-\EE\big[\sum\limits_{n=1}^N u_n(X^n+Y^n) \big]\big)^+, 
\end{align}
where $\mu, \lambda$ are hyperparameters and we write $$\tilde{\rho}(\mathbf{X}) = \inf_{w, b} J_{\text{primal}}(\varphi).$$

In \eqref{eq:rho_tilde}, the second term is the penalty for the variance and the third term is the penalty for failure of falling into the acceptance set. With the proper choice of $\mu$ and $\lambda$, the two penalties are very close to $0$ at optimal, which renders $\tilde{\rho}(\mathbf{X}) \approx \rho(\mathbf{X})$. In practice, we will compute the empirical estimation of $J_{\text{primal}}(\varphi)$ by using Monte Carlo algorithm to estimate the variance and expectation in \eqref{eq:rho_tilde}. The detailed description is provided in Algorithm \ref{algo:primal}.

\begin{algorithm}
\caption{Primal problem.}
\label{algo:primal}
\begin{algorithmic}
\REQUIRE Data $\{\mathbf{X}(\omega_i)\}_{i=1}^{\text{batch}}$, neural net $\varphi$, functions $\{u_n\}_{n=1}^N$ and hyperparameters $\mu, \lambda$, $B$, learning rate $\gamma$, Epochs

\FOR{$e=1$ to Epochs}
\STATE compute empirical estimation $\hat{J}$ of $J_{\text{primal}}(\varphi)$
\STATE compute gradients $\nabla_w\hat{J}$ and $\nabla_b\hat{J}$
\STATE update $\varphi$: $w=w-\gamma\nabla_w\hat{J}$ and $b=b-\gamma\nabla_b\hat{J}$
\ENDFOR
\STATE compute empirical estimation: $\hat{\rho}=\hat{J}_{\text{primal}}(\varphi)$ 
\ENSURE Updated neural net $\varphi$, $\hat{\rho}$
\end{algorithmic}
\end{algorithm}

\subsection{Dual Problem}
Without loss of generality, while the reason and discussion on general case will be postponed to Remark~\ref{remark:single_group}, we can just consider the single group case where all measures $Q^n$'s are the same by \cite{BFFM-fairness},  and we will simply write $Q$ instead of $Q^n$ for all $n$ in the following discussion. That is, the restriction in \eqref{eq:dual_restriction_on_Q} can be automatically ignored because we are able to interchange the order of expectation and summation. We estimate the measure $Q$ by using neural network $\Theta: \RR^N \to \RR^+$, parameterized by $(w_\theta, b_\theta)$, to represent its Radon-Nikodym derivative $\frac{d Q}{d \PP}$ with respect to the physical measure. $\Theta$ takes $\mathbf{X}$ as input and generates nonnegative output with unit mean, which can be realized by using a final \textbf{Softplus}\footnote{$\text{Softplus}(x) = \log(1+\exp(x))$.} layer and dividing the outputs by their average. Since $\Theta$ has $\omega$-by-$\omega$ output, we shall construct another neural network $\Psi:\RR^N \to \RR^N$, parameterized by $(w_\psi, b_\psi)$, taking $\mathbf{X}$ as arguments to generate random variables 
$$\mathbf{Z}:=(Z^1,\dots, Z^N)=(\Psi_1(\mathbf{X}), \dots, \Psi_N(\mathbf{X}))=:\Psi(\mathbf{X})$$ 
in evaluating $\alpha_B(\mathbf{Q})$. That is
$$\alpha_B(\mathbf{Q}) \equiv \alpha_B(\Psi, \Theta) = \sup_{w_\psi, b_\psi}\left\{ \sum_{n=1}^N \EE[-\Psi_n(\mathbf{X})\Theta(\mathbf{X}) ] \right\}.$$
Similar to the primal case, we have constraint imposed by the acceptance set $\mathcal{A}$ and thus include penalty when the random variable $\mathbf{Z}$ falls outside of $\mathcal{A}$. Denote by 
\begin{align*}
    J_\alpha (\Psi, \Theta)=  \sum_{n=1}^N \EE[-\Psi_n(\mathbf{X})\Theta(\mathbf{X}) ] - \lambda_\alpha \big(B - \sum_{n=1}^N \EE\big[ u_n(\Psi_n(\mathbf{X}))\big]\big)^+ 
\end{align*}
the objective function to be optimized for $\alpha_B(\mathbf{Q})$, and we write 
$$\tilde{\alpha}_B(\Psi, \Theta) = \sup_{w_\psi, b_\psi} J_\alpha (\Psi, \Theta).$$
Therefore, the objective function for dual representation is given by
\begin{align*}
    J_{\text{dual}}(\Psi, \Theta) := \sum_{n=1}^N \EE\big[ -X^n\cdot \Theta(\mathbf{X}) \big] - \tilde{\alpha}_B(\Psi, \Theta)
\end{align*}
and we write
$$\tilde{\rho}(\mathbf{X}) = \sup_{w_\theta, b_\theta} J_{\text{dual}}(\Psi, \Theta).$$
At optimal, $\tilde{\alpha}_B$ and $\tilde{\rho}$ can approximate real $\alpha_B$ and $\rho$ very well in the dual Problem \ref{problem:dual} since the penalty terms are almost $0$. Note that the first part in $J_{\text{dual}}$ does not depend on the parameters $w_\psi, b_\psi$ of neural network $\Psi$. Our model is trained in the same fashion of training GANs and we provide a detailed description\footnote{We use $\hat{\EE}$ to represent empirical expectation in this algorithm.} in Algorithm \ref{algo:dual}.

\begin{algorithm}
\caption{Dual problem.}
\label{algo:dual}
\begin{algorithmic}
\REQUIRE Data $\{\mathbf{X}(\omega_i)\}_{i=1}^{\text{batch}}$, neural nets $\Psi, \Theta$, functions $\{u_n\}_{n=1}^N$ and hyperparameters $\lambda_\alpha$, $B$, learning rates $\gamma_1, \gamma_2$, Epochs

\FOR{$e=1$ to Epochs}
\STATE compute empirical estimations $\hat{J}_\alpha$ of $J_{\alpha}(\Psi, \Theta)$ and $\hat{J}_{\text{dual}}$ of $J_{\text{dual}}(\Psi, \Theta)$ by
$$\hat{J}_{\text{dual}}=\sum_{n=1}^N \hat{\EE}\big[ -X^n\cdot \Theta(\mathbf{X}) \big] -\hat{J}_\alpha $$

\STATE compute gradients $\nabla_{w_\psi}\hat{J}_{\text{dual}}, \nabla_{b_\psi}\hat{J}_{\text{dual}}$ and $\nabla_{w_\theta}\hat{J}_{\text{dual}}, \nabla_{b_\theta}\hat{J}_{\text{dual}}$

\STATE update $\Psi$: $$w_\psi=w_\psi - \gamma_1\nabla_{w_\psi}\hat{J}_{\text{dual}},\quad b_\psi=b_\psi - \gamma_1\nabla_{b_\psi}\hat{J}_{\text{dual}}$$

\STATE update $\Theta$: $$w_\theta=w_\theta + \gamma_2\nabla_{w_\theta}\hat{J}_{\text{dual}}, \quad b_\theta=b_\theta + \gamma_2\nabla_{b_\theta}\hat{J}_{\text{dual}}$$

\ENDFOR
\STATE compute empirical: $\hat{\alpha}_B(\mathbf{Q})=\hat{J}_\alpha(\Psi, \Theta)$ and $\hat{\rho}=\hat{J}_{\text{dual}}(\Psi, \Theta)$
\ENSURE Neural nets $\Psi, \Theta, \hat{\alpha}_B(\mathbf{Q}), \hat{\rho}$
\end{algorithmic}
\end{algorithm}

\begin{remark} \label{remark:single_group}
For simplicity of description, we first consider only one group in an $N$-institution system and the optimal risk allocation strategy from the primal and dual problem. In this way, we are able to get rid of the restriction described in \eqref{eq:dual_restriction_on_Q} and it is very convenient to construct our algorithm.

In general case with $h$ non-overlapping groups for $N$ financial institutions, assuming the index sets are $I_1,\ldots,I_h$, the set of admissible random allocations should be $\mathcal{C}^d =\mathcal{C}_0^d\cap M^{\Phi} $ where
\[
\mathcal{C}_0^d = \left\{\mathbf{Y}\in L^0(\mathbb{R}^N)\,|\,\exists\,d_m\in \mathbb{R}:\,\sum\limits_{i \in I_m} Y^i = d_m,\,\forall\, m = 1,\ldots,h\right\}.
\]
It implies that the sum of components of  $\mathbf{Y}$ in each group must sum up to a real number, i.e. deterministic. The group sums $(d_1,\ldots,d_h)$ may change but the elements in each group are fixed. Thus $\mathcal{C}^d$ is a linear space containing $\mathbb{R}^N$ and the dual representation \ref{problem:dual} also applies. Biagini et al. \cite{BFFM-fairness} concludes that for institutions $i,j$ in the same group, the measures $Q^i,Q^j$ of the dual elements are the same. 
Thus in a multi-group system, we can use $h$ neural networks $(\Theta^1, \dots, \Theta^h)$ to represent $h$ different Radon-Nikodym derivatives. Then the objective function for $\alpha_B(\mathbf{Q})$ becomes
\begin{align*}
J_\alpha(\Psi, \Theta_1, \dots, \Theta_h):= & \sum_{j=1}^h \sum_{i\in I_j} \EE[-\Psi_i(\mathbf{X}) \Theta^j(\mathbf{X})] \nonumber \\
&\qquad - \lambda_\alpha \big(B - \sum_{n=1}^N \EE\big[ u_n(\Psi_n(\mathbf{X}))\big]\big)^+ ,
\end{align*}
and again, the restriction in $\eqref{eq:dual_restriction_on_Q}$ is satisfied automatically. The experiment of the general case is shown in section \ref{sec:Multi-groupExp} with $h=3$ and flowcharts of the algorithms are illustrated in Figure \ref{fig:primal-dual multigroup}.

\begin{figure}[h]
  \centering
  \includegraphics[width=0.8\linewidth]{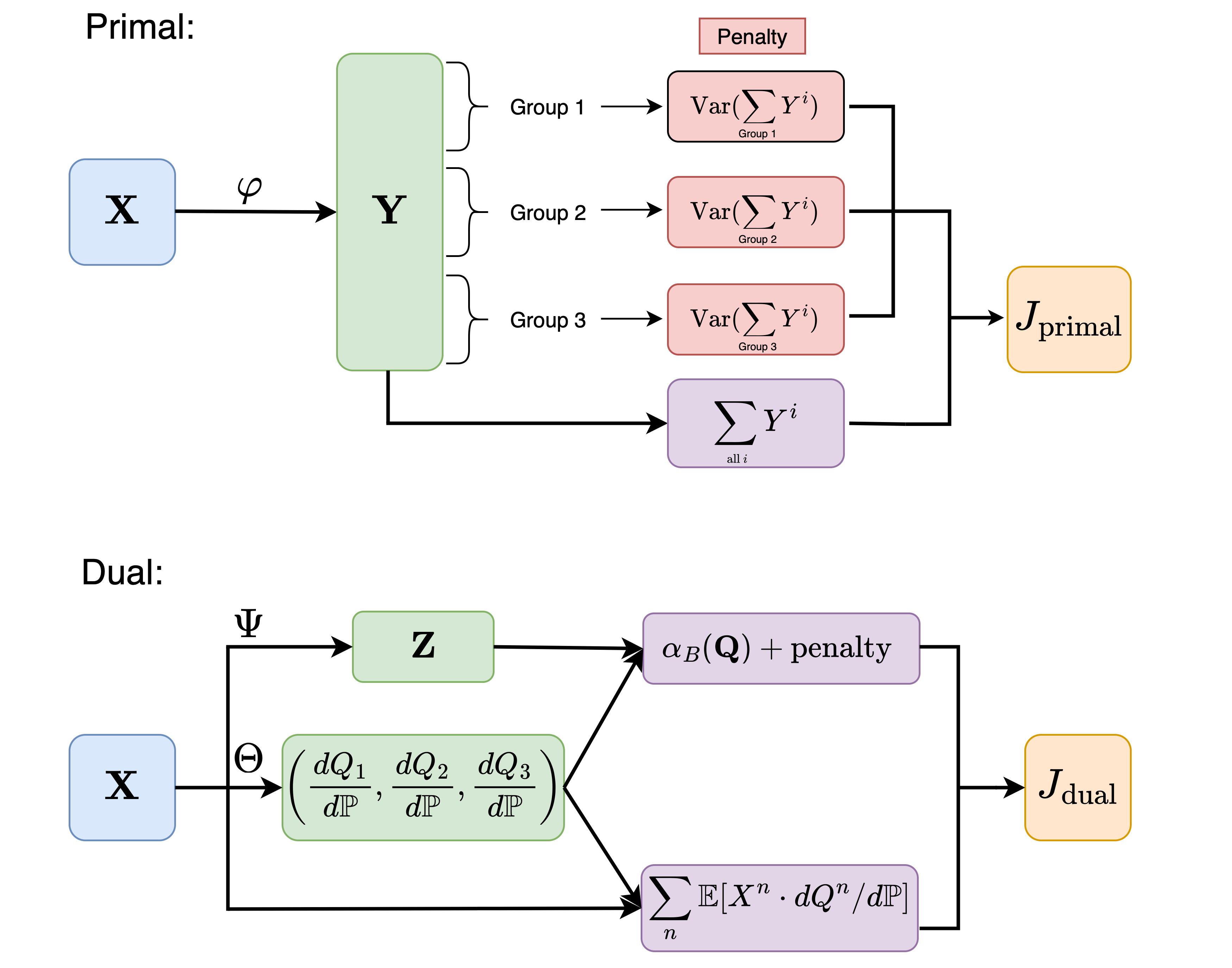}
  \caption{Illustration of multi-group algorithms.}
  \label{fig:primal-dual multigroup}
\end{figure}

\end{remark}

In this section, we construct algorithms that provide an approximation to risk allocations $\varphi(\mathbf{X}),\tilde{\rho}$ and Radon-Nikodym derivative $\Theta(\mathbf{X})$. Our algorithms are implemented in the context of the numerical examples that are presented in section \ref{section4:exp}. We conclude this section by providing the solution to \textit{fair} risk allocations \eqref{eqn:IndividualRiskAllocation} of each financial institution which combines Algorithm \ref{algo:primal}-\ref{algo:dual}, 

$$\rho^n(\mathbf{X})=\EE\big[ \varphi_n(\mathbf{X}) \Theta(\mathbf{X}) \big],\quad \forall n=1,\dots,N.$$

\section{Experiments}
\label{section4:exp}
We justify Algorithm \ref{algo:primal}-\ref{algo:dual} with two experiments. First, we employ exponential utilities for financial institutions with Gaussian distributed risk factors and compare experiment results with explicit results shown in \cite{BFFM-fairness}. Secondly, we apply the same setup to Problem \ref{problem:MoreRealisticSetup} and present numerical results as solutions. We conclude that given a financial system, we can predict the overall risk allocation and the distribution of risk allocations for all individual banks at terminal time $T$, as well as the \textit{fair} risk allocations for institutions at time $0$. The algorithms can be used to help decide, in order to secure a system in terms of the systemic risk measures defined in Problem \ref{problem:primal} or Problem \ref{problem:MoreRealisticSetup}, how much cash from each individual institution $n$ should be collected, i.e. $\EE_Q[Y^n]$, and how much to distribute, i.e. $Y^n(\omega)$ under the scenario $\omega$, at the beginning and end of the period, respectively.

Both of our training and testing data consist of 50000 samples. Each sample is a realization of the $N$ ($=10$) dimensional vector $\mathbf{X}(\omega)$, representing risk factors of $10$ positively correlated financial institutions who are jointly normally distributed. We choose exponential utility functions $u_n = -\frac{e^{-\alpha_n x}}{\alpha_n},\,\alpha_n>0$, $n=1,\ldots,N$. We vary the utility parameters for all financial institutions and we select $B < 0$. We use SGD as our deep learning optimizer for all experiments. 
To assess the quality of our algorithm, we consider the following evaluation metrics:
\begin{itemize}
	\item \textbf{Absolute difference.} Absolute value of the difference between estimation and theoretical solution.
    \item \textbf{Overall relative difference (ORD).} Let $\hat{E}$ be an estimation of $E$, we define the ORD by
    $$R(\hat{E}, E) = \frac{\|\hat{E} - E\|_1}{\|E\|_1}$$
    with $\|\cdot\|_1$ as the $l_1$ metric when $E$ is a deterministic vector, and $\|\cdot\|_1$ as the $L_1$ metric when $E$ is a random variable.
\end{itemize}
Small values of evaluation metrics imply better performance.

\subsection{Single-group Exponential Utility}
\label{sec:OnegroupExp}
We focus on the case under section 6 of \cite{BFFM-fairness}. Under the same assumptions, the optimal total risk allocation $\rho(\mathbf{X})$, the optimal penalty, the optimal allocation $\mathbf{Y}$ and the optimal measure optimizer $Q$ of the corresponding dual Problem \ref{problem:dual} are explicitly given by, for each scenario $\omega$,
\begin{align}
&\rho(\mathbf{X})= \beta\log\left(-\frac{\beta}{B}\mathbb{E}\left[\exp (-S/\beta)\right]\right),\label{eqn:optrho}
\\
&\alpha_B(Q) = \beta \mathbb{E}\left[\frac{dQ}{d\mathbb{P}}\,\ln\left(\frac{dQ}{d\mathbb{P}}\right)\right]+\beta\,\ln\left(-\frac{B}{\beta}\right),
\\
&Y^n(\omega) = -X^n(\omega)+\frac{1}{\beta\alpha_n}S(\omega)+\frac{1}{\beta\alpha_n}\rho,\label{eqn:optY}
\\
&\frac{dQ}{d\mathbb{P}}(\omega) = \dfrac{\exp\left(-S(\omega)/\beta\right)}{\mathbb{E}\left[\exp\left(-S/\beta\right)\right]},\label{eqn:optQ}
\end{align}
where $\beta = \sum_{n=1}^N\frac{1}{\alpha_n}$, and $S(\omega) = \sum_{n=1}^NX^n(\omega)$. And the optimal \textit{fair} individual risk allocation is given by
\[
\rho^n(\mathbf{X}) = \mathbb{E}_Q[Y^n] = \mathbb{E}\left[Y^n\cdot\frac{dQ}{d\mathbb{P}} \right] .
\]
We compare with the optimal solutions \eqref{eqn:optrho}-\eqref{eqn:optQ} generated by Monte Carlo method to show the accuracy of our proposed Algorithms \ref{algo:primal}-\ref{algo:dual}. 

\begin{figure}[h]
  \centering
  \includegraphics[width=0.7\linewidth]{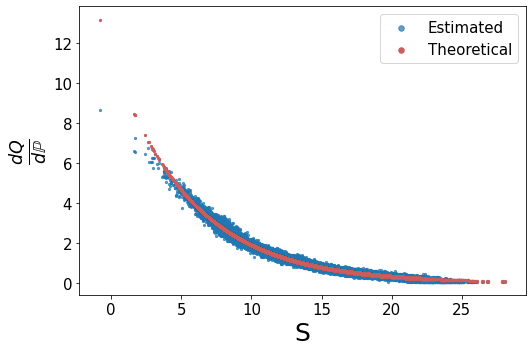}
  \caption{Behavior of $\widehat{\frac{dQ}{d\mathbb{P}}}(\omega)$ against $S(\omega)$.}
  \label{fig:Q-S}
\end{figure}

\paragraph{Evaluation.} First, to assess the goodness of estimation of Radon-Nikodym derivatives of the optimizer, we use overall relative difference to measure how one derivative function is different from the reference derivative function, i.e. 
\[
R(\widehat{\frac{dQ}{d\mathbb{P}}},{\frac{dQ}{d\mathbb{P}}}).
\]
It turns out the ORD of estimated Radon-Nikodym derivative is {\bf 4.55\%} and the behavior of the estimated measure derivative $\widehat{dQ/d\mathbb{P}}$ 
in terms of the sum of risk factors $S(\omega)$ for each scenario $\omega$ is shown in Figure~\ref{fig:Q-S}. The results both show it fits optimal Radon-Nikodym derivative very well.

We show performance of numerical results for the estimated overall risk allocation of ${\rho}$ and the estimated penalty of ${\alpha}$ in Table~\ref{tab:RhoAlpha}, along with the expected optimal results. The absolute differences (Abs. Difference) are quite small which indicates the estimation for the overall risk is successful.
\begin{table}
\centering
  \caption{Values of $\rho$ and $\alpha$}
  \label{tab:RhoAlpha}
  \begin{tabular}{ccl}
    \toprule
    &$\rho$ &$\alpha$ \\
    \midrule
    Estimated & -3.84 & -8.66\\
    Theoretical & -3.97 & -8.64\\
  \midrule
  \textbf{Abs. Difference}&0.13 &0.02
  \\
  \bottomrule
\end{tabular}
\end{table}

Table~\ref{tab:EQY} shows the estimated \textit{fair} risk allocations for all institutions
\[
\widehat{\mathbb{E}_Q[Y]}\,:=\, \left(\widehat{\mathbb{E}_Q[Y^n]}\right)_{n=1,\ldots,10} = \left(\mathbb{E}\left[\widehat{Y^n}\cdot \widehat{\frac{dQ}{d\mathbb{P}}}\right]\right)_{n=1,\ldots,10}
\]
and their theoretical optimal values. The overall relative difference (ORD) for them is defined as 
\[
R\,(\widehat{\mathbb{E}_Q[Y]},{\mathbb{E}_Q[Y]}).
\]
The ORD is {\bf 2.52\%} which shows great approximation of our algorithms to the theoretical optimal \textit{fair} allocation.
\begin{table*}
  \caption{Estimated and theoretical optimal values of ${\mathbb{E}_Q[Y^n]}$ in one-group.}
  \label{tab:EQY}
  \begin{tabular}{cccccccccccc}
    \toprule
     $n$&1 & 2 &3&4&5&6&7&8&9&10& \textbf{ORD}\\
    \midrule
     Estimated & 1.11 & 0.88 & 0.75 & 0.22 &-0.45 &-1.09 &-0.12 &-0.66 &-2.37 &-2.11& \multirow{2}{*}{2.52\%} \\
    Theoretical& 1.12 & 0.89 & 0.74  &0.18 &-0.47& -1.08& -0.16& -0.75 &-2.34 &-2.11
     \\
    \bottomrule
  \end{tabular}
\end{table*}
In conclusion, every optimal value we estimated shows excellent fitness with respect to the theoretical optimal value given by explicit formulas. Thus our algorithms provide reliable and accurate results.

\subsection{Multi-group Exponential Utility}
\label{sec:Multi-groupExp}
We verify our algorithm in the multi-group setup following the procedures described in Remark \ref{remark:single_group}. Similar to last section, the explicit solutions exist and the formulae are shown in detail in  \cite{BFFM-fairness}. 

\begin{figure*}[h]
  \centering
  \includegraphics[width=\linewidth]{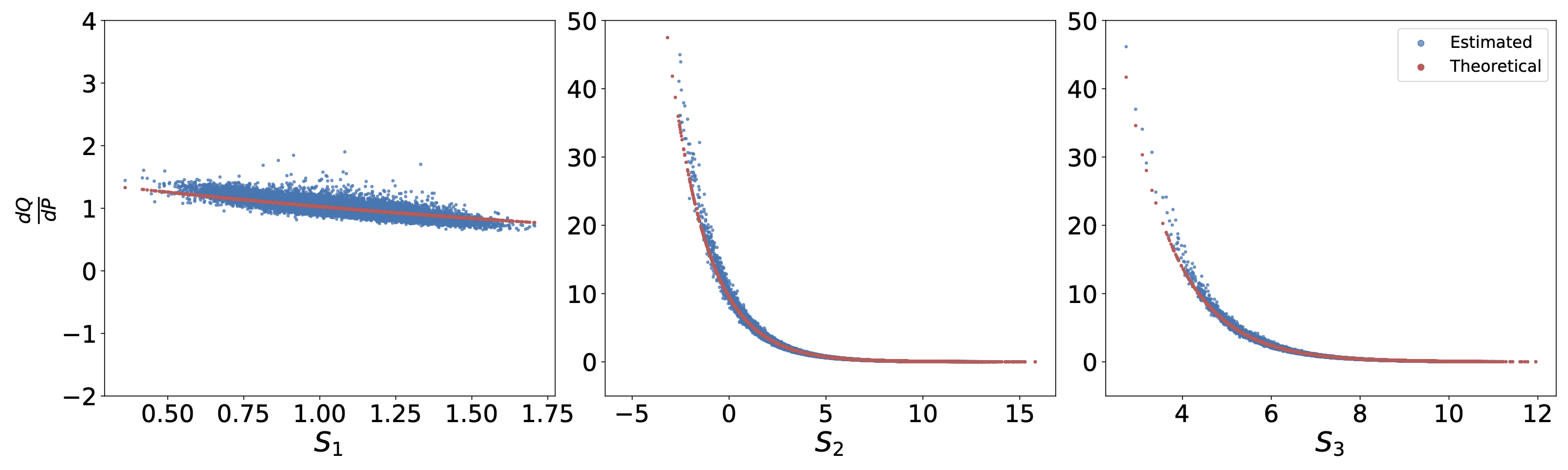}
  \caption{Behavior of $\widehat{\frac{dQ_i}{d\mathbb{P}}}(\omega)$ against $S_i(\omega)$, $i=1,2,3$.}
  \label{fig:Q-S_multi_group}
\end{figure*}

We consider separating $10$ institutions into $3$ groups with sizes $(3,4,3)$ and we will have one fair risk measure $Q$ for each group. By solving the dual problem, we get our estimated Radon-Nikodym derivatives for each $Q$ measure. Evaluating the estimated derivatives $\frac{d Q}{d \PP}$'s on testing data and comparing to the theoretical solutions, we have overall relative differences are $\mathbf{(3.42\%,6.31\%,5.12\%)}$ and we show the plots of behaviors of derivatives in terms of sums of group risk factors in Figure~\ref{fig:Q-S_multi_group}. The Radon-Nikodym derivative of each group is fitted very well and the distance between the estimated and the true functions is quite small. We then solve the primal problem to get estimated individual risk allocations $\hat{\mathbf{Y}}$, which renders us the estimate the fair risk allocations $\widehat{\EE_Q[\mathbf{Y}]}$. We compare our estimations with theoretical solutions in Table~\ref{tab:EQY_multi_group}. The estimated and optimal values in each pair are close to each other and the ORD is 3.9\% which demonstrates success of our algorithm in predicting individual fair systemic risk allocations in the multi-group case. Comparing with Table~\ref{tab:EQY}, the group sums of multi-group case are $(3.03,-0.92, -4.87)$ and the systemic risk measure is $-2.76$. The relative increase from $-3.84$ to $-2.76$ shows that the systemic risk measure increases when the system is separated into smaller subsystems, which is correct both intuitively and mathematically, since it is optimized over a smaller set in multi-group case.


\begin{table*}
  \caption{Estimated and theoretical optimal values of ${\mathbb{E}_Q[Y^n]}$ in multi-group.}
  \label{tab:EQY_multi_group}
  \begin{tabular}{cccccccccccc}
    \toprule
     $n$&1 & 2 &3&4&5&6&7&8&9&10&\textbf{ORD}\\
    \midrule
     Estimated & 1.19 & 1.11 & 0.73&  0.18& -0.55 &-1.41 & 0.86& 0.09 &-2.87& -2.09&\multirow{2}{*}{3.9\%}\\
    Theoretical& 1.25 & 1.16 & 0.77 & 0.17  & -0.56 & -1.42& 0.77 &  0.01 & -2.91 &-2.13
     \\
    \bottomrule
  \end{tabular}
\end{table*}

\subsection{No Cross-subsidization Measure}
Recall that in Problem \ref{problem:MoreRealisticSetup}, we proposed a the more realistic risk measure without cross-subsidization. We assume the risk allocations for all institutions are nonnegative which avoids withdrawing cash from some components. This is called no \textit{cross-subsidization} between financial institutions. However there is no explicit solution under this setup. So we will implement it with our algorithms and interpret the results with knowledge from experiment in section \ref{sec:OnegroupExp}. 

The estimated overall risk allocation
$
\hat{\rho} = 6.26;
$
and the estimated \textit{fair} risk allocations for all institutions are
\begin{align*}
&\left[\widehat{\mathbb{E}_Q[Y^1]},\ldots,\widehat{\mathbb{E}_Q[Y^{10}]}\right]
\\
    = &[0.69, 0.55, 0.43, 0.01, 0. , 0. ,  2.61 ,1.97 ,0. ,  0.  ].
\end{align*}
Comparing with the values in Table~\ref{tab:EQY} of section \ref{sec:OnegroupExp} which allows cross-subsidization, we can see the overall risk allocation for the system gets larger when we add constraints, which implies more cash is needed to secure the system. However the individual \textit{fair} risk allocation moves in no clear direction for every component, since it can be affected by many factors in the system, like individual risk factor distribution and correlations. Based on the fact that the utility parameters are in an increasing order in our experiment, we might reach a possible conclusion that more risk-averse institutions tend to contribute less when the system does not allow cross-subsidization. Future research will consist in studying the relation between risk allocation and factors by allowing control of parameters.

\subsection{Scalability}
Our algorithm is also scalable. We compare the training time for different size of financial system ($N$) and groups in Table~\ref{tab:time_scal}. Due to the fact all operations are realized via GPU in parallel, increasing the size of system only has subtle influence on training time. Increasing group size affects the calculation of our loss functions as well as the gradients of weights in neural networks, but the time increment is still small. 

\begin{table}[H]
\centering
  \caption{Training time.}
  \label{tab:time_scal}
  \begin{tabular}{ccccc}
    \toprule
     $(N, \text{groups})$ & $(10, 1)$ & $(10, 3)$ & $(100, 1)$ & $(100, 3)$\\
    \midrule
     Time (hours) & 0.38  & 0.43 & 0.42 & 0.47 \\
    \bottomrule
  \end{tabular}
\end{table}

\section{Conclusion}
In this work, we study systemic risk measures introduced in Problem \ref{problem:primal} and \ref{problem:MoreRealisticSetup}. Combined with dual problems, these measures can provide both the individual risk allocation ${Y}^n$ and the individual \textit{fair} risk allocation $\mathbb{E}_{Q^n}[Y^n]$ for each institution $n$. When applied to a real-life financial system that aims to secure the system from potential systemic risks, the latter can be set as a benchmark for collecting cash from institutions and the former can provide an accurate scenario-dependent estimation of cash distribution at the terminal time. In general there is no explicit solutions to these problems especially for the case where we propose a more realistic measure requiring no-subsidization between financial institutions. 
We show that the deep learning framework proposed in section \ref{section3:alg}, based on the delicately tailored objective functions and newly designed direct estimation of Radon-Nikodym derivative, yields excellent performance on estimating both the overall systemic risk allocation and individual \textit{fair} risk allocations. 
The algorithm is also efficient in two ways. Firstly in a financial system with a fixed number of institutions, solving the problem with more groups does not take more time than the problem with less groups. Secondly, when the dimension of the system being large ($N$), the processing time with our algorithm to compute optimal risk allocations has subtle difference. Moreover, the neural network based estimation of Radon-Nikodym derivative has broader applications in financial math, such as robust utility maximization and derivative pricing.

In summary, given the very large amount of institutions in the financial system, such techniques provide the suggestion/possibility of securing the system while keeping efficiency and therefore, mitigate the systemic risk faced by regulators nowadays.


\bibliographystyle{plain}
\bibliography{reference}

\end{document}